\documentclass[prb,twocolumn,showpacs,superscriptaddress,letterpaper]{revtex4}
\usepackage{graphicx}
\usepackage{amssymb}
\usepackage{citesort}
\usepackage{stmaryrd}
\begin{document}

\title{Theory of interfacial charge-transfer complex photophysics in $\pi$-conjugated polymer-fullerene blends.}
\author{K. Aryanpour}
\affiliation{Department of Physics, University of Arizona
Tucson, Arizona 85721, USA}
\author{D. Psiachos}
\affiliation{Department of Physics, University of Arizona
Tucson, Arizona 85721, USA}
\author{S. Mazumdar}
\affiliation{Department of Physics, University of Arizona
Tucson, Arizona 85721, USA}
\affiliation{College of Optical Sciences, University of Arizona
Tucson, Arizona 85721, USA}
\date{\today}
\begin{abstract}
We present a theory of the electronic structure and photophysics of 1:1 blends of derivatives of polyparaphenylenevinylene and fullerenes. Within the same Coulomb-correlated Hamiltonian applied previously to interacting chains of single-component $\pi$-conjugated polymers, we find an exciplex state that occurs below the polymer's optical exciton. Weak absorption from the ground state occurs to the exciplex. We explain transient photoinduced absorptions in the blend, observed for both above-gap and below-gap photoexcitations, within our theory. Photoinduced absorptions for above-gap photoexcitation are from the optical exciton as well as the exciplex, while for below-gap photoexcitation induced absorptions are from the exciplex alone. In neither case are free polarons generated in the time scale of the experiment. Importantly, the photophysics of films of single-component $\pi$-conjugated polymers and blends can both be understood by extending Mulliken's theory of ground-state charge transfer to the case of excited-state charge transfer.
\end{abstract}
\maketitle
\section{Introduction}
\label{sec:introduction}
\par In spite of intensive effort by many research groups over nearly two decades, the power-conversion efficiency of organic solar cells remains relatively low. \cite{Green08a} The most popular organic solar cells have been based on blends of $\pi$-conjugated polymers (PCPs) and fullerenes. \cite{Sariciftci92a,Brabec01a,Smilowitz93a,Morita92a,Li05a,Kim06a,Reyes05a,KKim07a,Parkinson08a,Peet07a,JKim07a} Within one-electron theory, \cite{Sariciftci92a} photoinduced charge transfer (PICT), which lies at the heart of organic photovoltaics, involves the photoexcitation of either the donor (D) PCP or the acceptor (A) fullerene, followed by a single-step electron (hole) transfer from the lowest antibonding (highest bonding) molecular orbital (MO) of the optically excited D (A) to the lowest available MO of A (D). Recent research by many groups have, however, detected an interfacial excited-state charge transfer complex (hereafter exciplex) below the optical gaps of both D and A upon photoexcitation. \cite{Veldman09a,Bruevich07a,Bakulin09a,Hallermann09a,Hwang08a,Westenhoff08a, Bakulin04a,Morteani04a,Sreearunothai06a,Goris06a,Osikowicz07a,Benson-Smith07a,Drori08a,Hallermann08a,Holt09a} It is now recognized that exciplex formation reduces the yield of mobile charge carriers. \cite{Veldman09a,Hwang08a,Westenhoff08a} Understanding the details of PICT is therefore clearly important for reaching enhanced device performance.
\par We present here a many-electron theory of PICT between derivatives of poly-paraphenylenevinylene (PPV) and C$_{60}$, focusing on the nature of the exciplex wave function, the difference between ground- and excited-state charge transfers and the photophysics of the exciplex. One goal of our work is to give consistent explanations of perplexing experiments in PCP-fullerene DA blends. A second broader goal is to show that there exists a universality in the photophsyics of films of single-component PCPs (Refs. 28 and 29) with no identifiable donor and acceptor constituents on the one hand, and blends on the other. As in the blends, there occurs an interfacial excited-state charge transfer complex below the strong optically allowed state also in the single-component systems. \cite{Rothberg06a,Yan94a,Hsu94a,Arkhipov04a,Schwartz03a,Ho01a,Lim02a,Brown03a,Koren03a} The complex that forms in films of single-component systems. \cite{Wang08a,Psiachos09a} is the excited-state equivalent of the ``neutral'' charge transfer complex of Mulliken. \cite{Mulliken52a} We show that the exciplex in blends is the excited-state equivalent of ``ionic'' charge transfer complexes. Differences in the photophysics of the two different kinds of intermolecular species can be rationalized within this context.
\par The bulk of the experimental works in blends have been limited to investigations of radiative and nonradiative couplings between the exciplex and the ground state, focusing on absorption, photoluminescence (PL) and electroluminescence (EL). In general, these studies indicate the occurrence of the exciplex below the optical states of both components, with weak radiative coupling to the ground state, but do not usually give the degree of charge transfer, hereafter ionicity, of the exciplex. Theoretical research on specific DA systems, where D and A are polyfluorene copolymers, has shown that the exciplex can have a varying degree of ionicity (including ionicity nearly 1) and has given satisfactory explanations of PL and EL from the exciplex. \cite{Huang08a} Much additional information is, however, obtained from ultrafast spectroscopy that detects photoinduced absorptions (PA). Ultrafast spectroscopy of films of single-component PCPs, for example, had early on found evidence for interchain species, \cite{Rothberg06a,Yan94a,Hsu94a} the nature of which has been of continuing interest.\cite{Wang08a} Remarkable advance in ultrafast spectroscopy of single-component PCPs was achieved recently by Sheng {\it et al.}, who extended the PA studies to the previously inaccessible mid infrared region, and in addition to the PA expected from the exciton (labeled PA$_1$ by the authors) found PA bands P$_1$ and P$_2$ at energies where charged polarons are expected to absorb. \cite{Sheng07a} Surprisingly, although this measurement suggested direct photogeneration of free polarons (in addition to excitons), infrared-active vibration (IRAV), known to be associated with free charges \cite{Ehrenfreund87a} was vanishingly weak. This apparent contradiction has been resolved in theoretical work that has shown that the PA in films of single-component PCPs originates from both the optical exciton and a bound charge transfer exciton (CTX) that occurs below the exciton. \cite{Wang08a,Psiachos09a} The CTX is a quantum-mechanical superposition of the intrachain exciton and the interchain Coulombically bound polaron pair, and exhibits picosecond (ps) PA at energies characteristic of both the neutral exciton and charged polarons. \cite{Psiachos09a} The absence of IRAV is due to the {\it symmetric} nature of the polaron-pair contribution to the CTX, with the probability of either chain being charged positively or negatively being exactly equal.
\par Drori {\it et al.} have recently performed ultrafast spectroscopy of a blend of MEH-PPV [where MEH stands for 2-methoxy-5-(2$^{\prime}$-ethylhexyloxy)] and C$_{60}$. \cite{Drori08a} The pump energy was set to both above and below the optical gap of the polymer (2.2 eV) at 3.1 and 1.55 eV, respectively (the lowest optically allowed state of C$_{60}$ occurs above the 1$B_u$ exciton of MEH-PPV). Above-gap photoexcitation shows at zero time delay all three PA bands seen in films of MEH-PPV, \cite{Sheng07a} viz., P$_1$, PA$_1$, and P$_2$, with peaks at $\sim$ 0.4, 1, and 1.6 eV, respectively. \cite{Drori08a} Unlike in pure MEH-PPV films, \cite{Sheng07a} however, PA$_1$ now decays rapidly (it nearly vanishes in 10 ps), and there is simultaneous growth in P$_1$ and P$_2$ intensities. In what follows, we denote excited-states with excitations predominantly on the polymer as MEH-PPV$^*$C$_{60}$. States with excitations predominantly on the C$_{60}$ are written as MEH-PPV-C$_{60f}^*$ or MEH-PPV-C$_{60a}^*$, where the subscripts $f$ and $a$ refer to dipole forbidden (HOMO$\to$LUMO transition in C$_{60}$ within one-electron theory) and dipole-allowed (HOMO$\to$LUMO+1 and HOMO-1$\to$LUMO) transitions, respectively. We use these notations for simple classification only, although the lower symmetry in the blend, compared to isolated C$_{60}$ molecule, implies that the MEH-PPV-C$_{60f}^*$ are not strictly forbidden optically. In principle, the time-dependent decay of PA$_1$ and growths of P$_1$ and P$_2$ can be due to dissociation of the MEH-PPV$^*$C$_{60}$ exciton into charged MEH-PPV$^+$ and C$_{60}^-$. PA from the MEH-PPV$^+$ component would then give both P$_1$ and P$_2$ (the opposite process, photogeneration of MEH-PPV-C$_{60a}^*$, followed by charge transfer from the HOMO of PPV to the HOMO of C$_{60}$ to generate MEH-PPV$^+$ and C$_{60}^-$ is not a realistic possibility, since this would not give the PA$_1$ band in transient absorption.) Surprisingly, Drori {\it et al.} find P$_1$ and P$_2$ in ultrafast measurements of PPV-C$_{60}$ even upon exciting at 1.55 eV, 0.6 eV {\it below} the MEH-PPV$^*$ exciton. This result is peculiar to the blend and is not observed in pure MEH-PPV films. In  contrast to pure MEH-PPV, IRAV signatures accompanying P$_1$ and P$_2$ are now strong, suggesting the possibility that free polarons are indeed generated directly. \cite{Drori08a} P$_1$ and P$_2$ are seen also in continuous-wave (cw) measurements, with both above- and below-gap excitations. The similarity between the below-gap and above-gap spectra (see Fig.~2 of Ref. 22) suggests free polaron generation even upon below-gap cw excitation. The polarons generated by below-gap cw excitation, however, have lifetimes two orders of magnitude larger than those generated by above-gap excitations and are far less mobile. \cite{Drori08a}
\par As in the case of single-component PCPs, \cite{Psiachos09a} P$_1$ and P$_2$ can originate from a bound state below the optical exciton instead of free polarons. The disappearance of the PA$_1$ band with simultaneous growth in P$_1$ and P$_2$ intensities, however, appears to support the original one-electron picture of exciton dissociation into free polarons. \cite{Sariciftci92a,Drori08a} Recall, for example, that PA at energy nearly the same as PA$_1$ occurs from the CTX in single-component systems. \cite{Psiachos09a} In the following, we arrive at consistent explanations of the above apparently contradictory experimental observations, from calculations within the same theoretical model that was previously used for MEH-PPV films alone. \cite{Wang08a,Psiachos09a} We explain the instantaneous generation of PA bands P$_1$ and P$_2$ by both above-gap and below-gap excitations, as well as the enormously long lifetime of the polarons generated by below-gap cw excitation and their immobility. We also explain the absence of the PA$_1$ band following below-gap excitation and give a plausible explanation for the rapid decay of the same upon above-gap excitation. 

\section{Theoretical model and method}
\label{sec:theory}
\par
\begin{figure}
\includegraphics[width=3.3in]{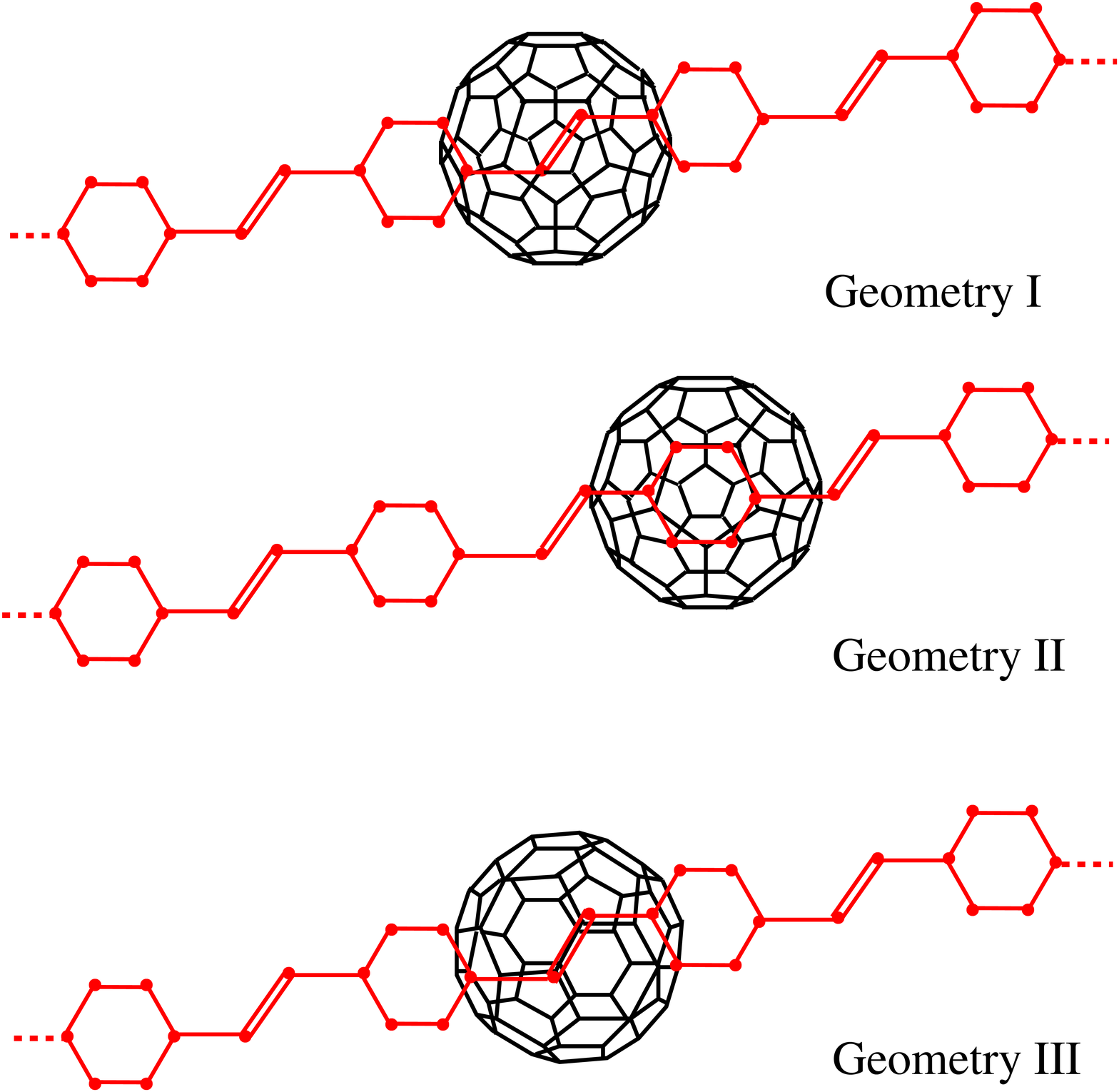}
\caption[a]{(Color online) Top view of the relative locations of the 8-unit PPV oligomer and the C$_{60}$ molecule assumed in our calculations for three different geometries I, II, and III. In geometry I, the topmost pentagon of C$_{60}$ is parallel to PPV plane and located symmetrically below an ethylenic bond of the 8-unit PPV oligomer. In geometry II, the same pentagon of C$_{60}$ is located symmetrically below a phenyl ring of the PPV oligomer. In geometry III, the central ethylenic bond of the PPV oligomer is parallel to the common bond between two hexagons in C$_{60}$.}
\label{blend}
\end{figure}

 We extend our previous Pariser-Parr-Pople (PPP) calculations \cite{Wang08a,Psiachos09a} to an idealized 1:1 PPV-C$_{60}$ blend, which we approximate by a planar capped PPV oligomer 8 units long, and a C$_{60}$ molecule whose nearest carbon atoms are located 0.4 nm below the plane of the PPV. We have considered three different geometries, with (i) a pentagonal face of the C$_{60}$ molecule parallel to the PPV plane and located symmetrically below the central ethylenic bond (see Fig.\ref{blend}, geometry I), (ii) the same pentagonal face now positioned below and parallel to a phenyl ring in the PPV oligomer (see Fig.\ref{blend}, geometry II), and (iii) a symmetric configuration with the ethylenic bond in the center of the 8-unit PPV occurring above the bond common to two hexagons in the C$_{60}$ molecule (see Fig.\ref{blend}, geometry III). The parallel orientation of a face or bond of C$_{60}$ with the plane of PPV is necessary to obtain the significant intermolecular charge transfer that is implied by experiments, viz., emission from the exciplex \cite{Benson-Smith07a,Hallermann08a} and PA even upon below-gap excitation. \cite{Drori08a} The three geometrical arrangements chosen, however, give large variations in intermolecular Coulomb interactions and electron hoppings (see below). We have confirmed that simple translational shifts of the C$_{60}$ molecule along the length of the PPV chain while maintaining the above relative orientations makes negligible differences.  
\par The overall Hamiltonian $H$ is written as $H = H_{intra}+H_{inter}$, where the intramolecular and intermolecular components are
\begin{eqnarray}
\label{H_intra}
H_{intra}=-\sum_{\mu\langle ij \rangle, \sigma}t_{ij}
(c_{\mu,i,\sigma}^\dagger c_{\mu,j,\sigma}+ \textrm{H.C.}) &+& \nonumber \\ 
U\sum_{\mu,i} n_{\mu,i,\uparrow} n_{\mu,i,\downarrow}
 + \sum_{\mu,i<j} V_{ij} (n_{\mu,i}-1)(n_{\mu,j}-1)
\end{eqnarray}
and
\begin{eqnarray}
\label{H_inter}
H_{inter}=-\sum_{\mu <\mu^{\prime},i,j,\sigma}t^{\perp}_{ij}(c^{\dagger}_{\mu,i, \sigma}
c_{\mu^{\prime},j,\sigma} + \textrm{H.C.}) \nonumber &+& \\
\frac{1}{2}\sum_{\mu < \mu^{\prime},i,j} V_{ij}^{\perp}(n_{\mu,i} -1)(n_{\mu^{\prime},j} - 1)\,.
\end{eqnarray}
\par In the above $c^{\dagger}_{\mu,i,\sigma}$ creates a $\pi$ electron of spin $\sigma$ on carbon atom $i$ of molecule $\mu (=1,2)$, with $\mu=1$ and 2 corresponding to PPV and C$_{60}$, respectively; $n_{\mu,i,\sigma} = c^{\dagger}_{\mu,i,\sigma}c_{\mu,i,\sigma}$ is the number of electrons on atom $i$ of molecule $\mu$ with spin $\sigma$ and $n_{\mu,i} = \sum_{\sigma}n_{\mu,i,\sigma}$ is the total number of electrons on atom $i$ of the molecule. For the PPV oligomer, we consider standard nearest-neighbor one-electron hopping integrals $t_{ij}$ = 2.4 eV for phenyl C-C bonds, and 2.2 (2.6) eV for single (double) C-C bonds, respectively. \cite{Chandross97a} We choose $t_{ij}$ = 1.96 eV for bonds within pentagons and $t_{ij}$ = 2.07 eV for bonds connecting the pentagons in C$_{60}$. The smaller $t_{ij}$ for C$_{60}$ reflect the curvature in fullerenes that reduces the $\pi$ overlap between neighboring $p$-orbitals. \cite{Wang06a} $U$ and $V_{ij}$ are the on-site and intrachain intersite Coulomb interactions, respectively. $V_{ij}$ are obtained by modifying the Ohno parametrization \cite{Ohno64a}
\begin{equation}
\label{Vij}
V_{ij}=\frac{U}{\kappa\sqrt{1+0.6117 R_{ij}^2}},
\end{equation}
where $R_{ij}$ is the distance between carbon atoms $i$ and $j$ in \AA, and $\kappa$ models the effective charge seen at a distance along the chain. \cite{Chandross97a} ($\kappa=1$ within the original Ohno model.)
\par We choose the same functional form for $V^\perp_{ij}$ as in Eq.~\ref{Vij}, with a variable intermolecular decay parameter $\kappa_{\perp} \leq \kappa$. For $t^\perp_{ij}$ we use the distance-dependent exponential form that has been used for the intertube hopping integral in double-walled carbon nanotubes.\cite{Uryu04} The expression is
\begin{equation}
\label{t_perp}
t^\perp_{ij}=\beta\exp[(c-d_{ij})/\delta],
\end{equation}
where the prefactor $\beta=0.2~\textrm eV$, $c$ is the minimum vertical distance between the molecules (0.4 nm here), $d_{ij}$ is the distance between atoms $i$ (belonging to C$_{60}$) and $j$ (belonging to the PPV oligomer), and the decay constant $\delta=0.045~\textrm {nm}$. We use a basis of localized Hartree-Fock MOs of the individual molecular units in our calculations. Eigenstates within this basis are superpositions of configurations with distinct intramolecular or intermolecular excitations, thereby making precise computations of charges on the PPV and C$_{60}$ components feasible. \cite{Wang08a} In our calculations reported below, we refer to the positive (negative) charge in the PPV (C$_{60}$) component of each eigenstate as the ionicity characterizing the eigenstate. 
\par Our calculations for the PPV-C$_{60}$ blend are within the single configuration-interaction (SCI) approximation, which retains the Hamiltonian matrix elements between all one-electron one-hole (1e-1h) excitations from the Hartree-Fock ground state. This is in contrast to our earlier quadruple CI (QCI) and multiple reference single and double CI (MRSDCI) calculations for pairs of PPV oligomers of lengths 3 and 4 units each (PPV3 and PPV4), respectively. \cite{Psiachos09a}While the QCI (MRSDCI) is a far more sophisticated many-body approach than the SCI, it retains all (dominant) triple and quadruple Hartree-Fock excitations, and hence the total number of configurations increases as a high power of the number of carbon atoms $N$. The total number of configurations retained in the QCI calculations for PPV3 oligomers, for instance, is 1833276. The large $N$ in the present case (122 instead of 60 in our previous work) alone would make such many-body calculations impractical. Additionally, the high frontier orbital degeneracies in C$_{60}$ (fivefold degeneracy of HOMO and HOMO-1 each and threefold degeneracy of LUMO and LUMO+1 each. See Fig.\ref{HFMO}) imply an enormously large number of quadruple excitations with large spin degeneracies, each of nearly equal importance. Taken together, the large $N$ and the degeneracies indicate that higher-order CI can be performed here only by imposing severe cutoffs in the number of excited configurations that are retained, which in turn would lead to strong loss of precision in our computational results. 
\par We overcome these problems by performing QCI calculations for a {\it different} DA system with smaller $N$ and no frontier orbital degeneracies: a pair of PPV3 oligomers, with site energies (electronegativities) of opposite signs for the carbon atoms on the different molecules. We report these calculations in the Appendix, where we show that this artificial DA complex reproduces the frontier orbital offsets of PPV-C$_{60}$ with suitable choice of the carbon atom site energies, and mimics the electronic structure and photophysics of PPV-C$_{60}$. Direct comparisons between our SCI calculations for PPV-C$_{60}$ on the one hand, and SCI as well as QCI calculations for an analogous system on the other hand thereby becomes possible, 
enabling us to make semiquantitative predictions for PPV-C$_{60}$. 
\section{Computational Results}
\label{sec:results}
\par
\begin{figure}
\includegraphics[width=2.0in]{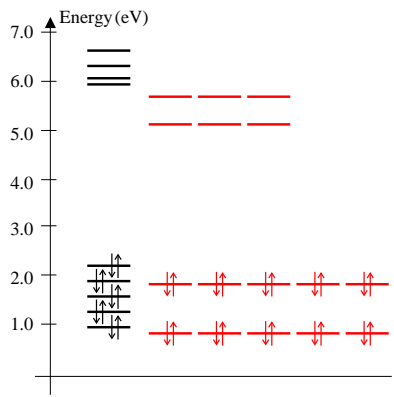}
\caption[a]{(Color online) The highest bonding and lowest antibonding Hartree-Fock MOs of the 8-unit PPV oligomer (left) and C$_{60}$ (right) within $H_{intra}$. The HOMOs of PPV and C$_{60}$ are at 2.18 and 1.80 eV, respectively. The corresponding LUMOs are at 5.82 and 5.18 eV, respectively.}
\label{HFMO}
\end{figure}

 In Fig.~\ref{HFMO} we have given the mean-field Hartree-Fock energy levels for the frontier MOs of PPV and C$_{60}$ in the limit of $H_{inter}=0$. The offsets between the HOMOs and the LUMOs identify PPV$^*$ as a donor and C$_{60}$ as a strong acceptor, with PPV-C$_{60}$ forming what has been termed a type-II heterostructure. The Hartree-Fock energy difference between PPV$^*$C$_{60}$ and PPVC$_{60f}^*$ is 0.64 eV, comparable to the SCI exciton-binding energy in PPV. \cite{Chandross97a} Based on the energy offsets alone therefore it is not possible to predict the ionicity of any exciplex formed.
\par Our motivation for considering the three different geometries in Fig.~\ref{blend} was to determine whether there occurs any significant difference between various relative orientations of the PPV oligomer and C$_{60}$. In reality, in view of the relatively few and small $t_{ij}^{\perp}$ that are possible 
between PPV and C$_{60}$, and $V_{ij}^{\perp}$ being dependent only on intermolecular distance rather than orientation, we expect, and find, small difference between the three cases. We therefore present detailed results only for geometry I in Fig.\ref{blend}. We follow this up with a comparison of the three different cases, where we show that the energies the calculated PA bands remain virtually the same for all orientations, although the strengths of the absorptions do differ moderately.
\par
\begin{figure}
\includegraphics[width=3.0in]{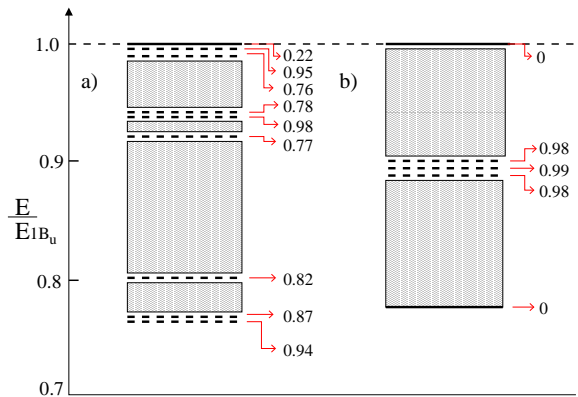}
\caption[a]{(Color online) Below-gap electronic structure of the PPV-C$_{60}$ blend for (a) $\kappa_{\perp}=1.3$ and (b) $\kappa_{\perp}=2.0$ (geometry I). The dashed lines correspond to exciplexes with the numbers against them giving their ionicities. The hatched regions are occupied by fifteen discrete PPV-C$_{60f}^*$ states with tiny energy gaps between them.}
\label{eng-spctrm}
\end{figure}

 In Figs.~\ref{eng-spctrm}(a) and \ref{eng-spctrm}(b) we show the results of SCI calculations for two representative $\kappa_{\perp}$ for geometry I. The reasons for performing calculations for $\kappa_{\perp}<\kappa$ are given below. The calculated energies of the PPV$^*$C$_{60}$ optical exciton for $\kappa_{\perp}=1.3$ and 2.0 are nearly the same, viz., 2.77 and 2.76 eV, respectively. The calculated SCI energy $E_{1B_u}$ of the 1$B_u$ optical exciton of the isolated PPV chain within the same $H_{intra}$ is $2.76~\textrm eV$. \cite{Chandross97a} Here and in subsequent figures, we scale all energies with respect to the calculated $E_{1B_u}$. We obtain estimates of the true energies of the different excited-states by simply multiplying the scaled energies by the experimental $E_{1B_u}$ = 2.2 eV. The purpose of this is to correct for the quantitative inaccuracies associated with SCI and to arrive at theoretical estimates that can then be compared against experiments. \cite{Wang08a,Psiachos09a} We have shown in Fig.~\ref{eng-spctrm} all energy states below the optical PPV$^*$C$_{60}$ exciton, including the charged exciplex states and the neutral PPVC$_{60f}^*$ states. SCI removes the degeneracy of the latter, and the fifteen nondegenerate states are now spread out over a narrow energy region of width $\sim$ 0.5 eV. Because of the very small energy gaps between these states, in Fig.~\ref{eng-spctrm} and elsewhere below we refer to these collectively as a ``band'' of states. The appearance of these states below the allowed exciton in our calculations is in agreement with the observation that absorption in the blend is to the PPV$^*$C$_{60}$ exciton, but emission occurs from the lower energy forbidden  PPVC$_{60f}^*$ states. \cite{Benson-Smith07a,Hallermann08a} The width of the experimental emission band from a blend of a PPV derivative and the fullerene PCBM ([6,6]-phenyl C$_{61}$-butyric acid methyl ester) \cite{Hallermann08a} is comparable to the scaled width of the band of forbidden states in Fig.\ref{eng-spctrm} ($\sim$ 0.5 eV).
\par In contrast to the weak effect $\kappa_{\perp}$ has on the energy of the PPV$^*$C$_{60}$ exciton, its effect on the relative energies of PPVC$_{60f}^*$ and the exciplex states is much stronger. The three nearly degenerate exciplex states with ionicities close to 1 in Fig.~\ref{eng-spctrm} (b) involve complete charge transfer between the HOMO of the PPV oligomer and the three degenerate LUMOs of the C$_{60}$ molecule. Our motivation for performing calculations with smaller $\kappa_{\perp}$ arises from the experiments that indicate emission \cite{Benson-Smith07a,Hallermann08a} and PA (Ref. 22) from the exciplex. The exciplex emission from the same PPV derivative and PCBM mentioned above \cite{Hallermann08a} occurs {\it below}  the broad emission from the forbidden fullerene states, exhibiting a peak at $\sim$ 1.48 eV. This energy is close to the energy of the lowest exciplex estimated from ultrafast spectroscopy ($\sim$ 1.55 eV). \cite{Drori08a} Both experiments therefore indicate that $\kappa_{\perp}=2$ does not reproduce the experimental energy location of the lowest exciplex. Our choice of $\kappa_{\perp}$ is phenomenological, based on the observation that the exciplex occurs below the ``band'' of PPVC$_{60f}^*$ states. For geometry I, $\kappa_{\perp}=1.3$ is the largest intermolecular dielectric constant for which this requirement is met. Our estimate for the energy of the exciplex for $\kappa_{\perp}=1.3$, using the above scaling procedure is 1.7 eV, close to the experimental estimates. \cite{Hallermann08a,Drori08a} Furthermore, a large number of excited exciplex states (with decreasing ionicities) occur in this case below the optical exciton, also in agreement to what has been inferred from experiments. \cite{Drori08a} It thus appears that  $\kappa_{\perp}=1.3$ rather than $\kappa_{\perp}=2.0$ 
gives a better fit to the experimental systems.
\par
\begin{figure}
\includegraphics[width=3.0in]{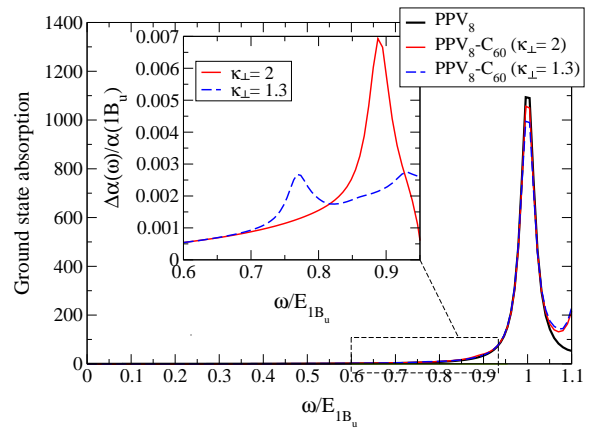}
\caption[a]{(Color online) Main panel: optical absorption to the lowest exciton in pure PPV and PPV-C$_{60}$ blend with $\kappa_{\perp}=1.3$ and $2.0$ (geometry I). The inset shows the subgap difference absorptions, normalized to the absorption at the peak of the exciton for both $\kappa_{\perp}$.}
\label{GS-absorption}
\end{figure}

 In Fig.~\ref{GS-absorption} we show the calculated ground state absorptions for the single chain PPV oligomer and the PPV-C$_{60}$ blend (geometry I). As noted experimentally, there is no perceptible difference in absorption studies on the scale of the figure. \cite{Hallermann08a} Weak absorptions below the exciton have been detected using Fourier transform photocurrent spectroscopy \cite{Goris06a} and photothermal deflection spectroscopy. \cite{Benson-Smith07a} The inset of the figure shows our plot of $\Delta \alpha(\omega)/\alpha(1B_u)$, where $\alpha(1B_u)$ is the absorption at the frequency of the 1$B_u$ exciton of the isolated PPV chain, and $\Delta \alpha(\omega)$ is the frequency-dependent difference in absorption between the blend and the single chain. Positive $\Delta \alpha$ in the subgap region indicates absorption to the exciplexes. The scaled lowest-calculated exciplex energy (1.7 eV) for $\kappa_{\perp}=1.3$ is close to the energy where weak absorption is detected experimentally, $\sim$ 1.6 eV. \cite{Goris06a,Benson-Smith07a} Our calculated $\Delta \alpha(\omega)/\alpha(1B_u)$ is, however, smaller by nearly an order of magnitude. The larger experimental subgap absorptions may be due to both disorder and electron-phonon interactions.
\par
\begin{figure}
\includegraphics[width=3.2in]{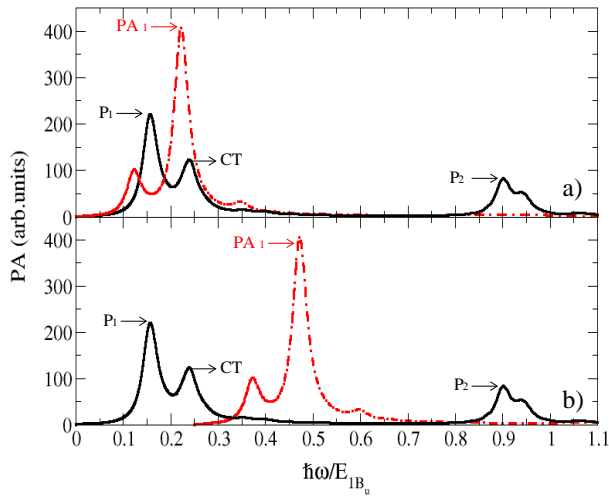}
\caption[a]{(Color online) Calculated PAs for photoexcitations above (red dot-dashed) and below (black solid) the optical exciton PPV$^*$-C$_{60}$ for $\kappa_{\perp}=1.3$ (geometry I). In panel (b), the calculated PA$_1$ has been rigidly blue-shifted to its experimental location.}
\label{PA-absorption}
\end{figure}

 In Fig.~\ref{PA-absorption}(a) we have plotted our calculated excited-state absorptions for $\kappa_{\perp}=1.3$ for comparisons to above- and below-gap transient PA bands, \cite{Drori08a} respectively (geometry I). Absorption labeled PA$_1$ is from the PPV$^*$C$_{60}$ optical exciton, while absorptions labeled P$_1$, CT, and P$_2$ are from the lowest exciplex. The calculated energy of PA$_1$ is considerably lower than what is found experimentally. Our scaling procedure gives the calculated PA$_1$ energy to be $\sim$ 0.5 eV, in contrast to the experimental PA$_1$ peak energy of $\sim$ 1.0 eV. \cite{Drori08a} PA$_1$ is predominantly the transition from the PPV$^*$C$_{60}$ exciton to the mA$_g$ two-photon state of PPV that dominates its optical nonlinearity. \cite{Chandross97a} The weak calculated PA on the low energy side of PA$_1$ has a charge transfer origin that results from weakly ionic character of the optical exciton for $\kappa_{\perp}=1.3$. It is known that CI with two-electron two-hole excitations lowers the energy of the 2A$_g$ and raises the energy of the mA$_g$. \cite{McWilliams91a} On the other hand, P$_1$ and P$_2$ absorptions are predominantly intraband 1e-1h excitations (see below) that should be weakly affected by higher-order correlations. Similar differences in extent of correlation effects also characterize one- and two-photon excitations from the ground state of neutral chains. It is thus to be expected that higher-order CI calculations in the present case should increase PA$_1$ energy, leaving P$_1$ and P$_2$ energies almost intact. We demonstrate this in the Appendix, where we compare SCI and QCI calculations for our model DA system mentioned before, pairs of PPV3 oligomers with carbon atom-site energies with opposite signs. As shown there, the SCI and QCI energies of the PA bands originating from the exciplex are only marginally different, but QCI blue-shifts the energy of the PA$_1$ band by a large amount. Based on this result, in order to simulate the higher-order CI effects we have therefore rigidly shifted the PA$_1$ peak to the energy expected from experiments \cite{Drori08a} in Fig.~\ref{PA-absorption}(b). 
\par We discuss the origin and the natures of PAs labeled P$_1$, P$_2$, and CT in the figure separately below. We have done detailed wave-function analyses of the final states reached in each of these absorptions. The large ionicity of the lowest exciplex, 0.94 (see Fig.~\ref{eng-spctrm}), implies that PAs from this state are close to polaron absorption energies of the PPV oligomer. These are the absorptions labeled P$_1$ and P$_2$. The P$_1$ absorption is predominantly a superposition of intramolecular ``intraband'' excitations from the ground state of the PPV$^+$ cation. The dominant excitation that describes this transition is from the HOMO-1 to the HOMO of the PPV oligomer. Several other excitations from lower-energy bonding MOs to the HOMO also contribute to the transition, as is expected for the polaron with interacting electrons. There occur, however, {\it additional} contributions to the P$_1$ transition beyond these purely intramolecular polaronic ones. The ionicity of the final state of the P$_1$ absorption, 0.78, is less than that of the initial exciplex. The smaller ionicity implies greater mixing between PPV$^+$C$_{60}^-$ and neutral PPVC$_{60a}^*$ in the final state of the P$_1$ absorption than in the initial state. We have confirmed this from wave-function analysis. Interestingly, there is almost no mixing with the forbidden PPVC$_{60f}^*$ excitations. P$_1$ therefore has a weak ``back charge transfer'' component.  
\par The dominant contribution to the P$_2$ absorption is from the singly occupied HOMO to the LUMO of PPV in the exciplex, in addition to higher-energy intramolecular excitations. Because of the rigid-band approximation adopted in our work, P$_2$ is nearly at the same energy as ground state absorption. The ionicity of the final state in this case is 0.87, only weakly different from that of the exciplex. The contribution of back charge transfer components to this transition is therefore smaller than in P$_1$.
\par The final state of the PA labeled CT has an ionicity 0.68, considerably smaller than that of the initial exciplex as well as the final states of P$_1$ and P$_2$. The contribution of PPVC$_{60a}^*$ configurations to the final state here are therefore much larger than in P$_1$ and P$_2$, and this is why we have labeled this additional PA band as CT. This PA band has so far not been seen experimentally in the PPV-C$_{60}$ blend, but has been observed  elsewhere, as we point out in the next section. A similar intermolecular charge transfer PA band has been seen in our previous calculations for coupled PPV chains, where, however, the CT band appears on the low-energy side of P$_1$. \cite{Psiachos09a} In both cases the CT band is expected from Mulliken's theory of charge transfer. \cite{Mulliken52a} Within Mulliken's theory, an interunit charge transfer absorption is expected independent of whether the lower state is neutral or ionic, with only the direction of charge transfer different. In the case of coupled PPV chains, the CT absorption is from the neutral CTX to the ionic polaron-pair state. \cite{Wang08a,Psiachos09a} Here the CT absorption is from the ionic PPV$^+$C$_{60}^-$ to the neutral PPVC$_{60a}^*$ states. The larger CT energy in the present case is a consequence of the HOMO and LUMO offsets in Fig.~2. P$_1$, P$_2$, and CT bands also have contributions from C$_{60}^-$ excitations.
\begin{figure}
\includegraphics[width=3.2in]{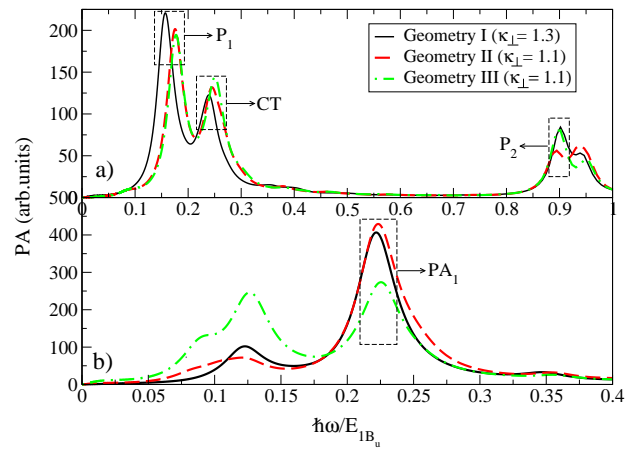}
\caption[a]{(Color online) Calculated PAs originating from (a) the exciplex, and (b) the exciton for the three different geometries in Fig.\ref{blend}. Adjustments of $\kappa_{\perp}$ values are done to achieve ionicity of $\sim$ 0.95 for the exciplex in PPV-C$_{60}$.}
\label{geom-comparison}
\end{figure}
\par We now make comparisons of our calculations based on the three different geometries of Fig.\ref{blend}. Geometries II and III give overall $t^{\perp}_{ij}$ that are slightly larger than those obtained with geometry I. This increases the degree of charge transfer, which in turn makes the exciplex less ionic. In order to make the exciplex as ionic as with geometry I, and the absolute lowest-excited state as suggested from experiments (see above), a larger $V_{ij}^{\perp}$, and therefore smaller $\kappa_{\perp}$ become necessary. With geometries II and III, this is achieved with $\kappa_{\perp}=1.1$, only slightly different from the $\kappa_{\perp}=1.3$ needed with geometry I. As seen in Fig.\ref{geom-comparison}, the energy locations of the PA bands from the exciton and the exciplex are very weakly affected by the geometry change. The effect on the intensities of the absorptions, in particular, the absorptions from the exciton, is stronger. This is to be expected from the large intermolecular hoppings in geometries II and III.
\section{Interpretation of experiments and discussions.}
\label{sec:discussion}
\par
\begin{figure}
\includegraphics[width=3.0in]{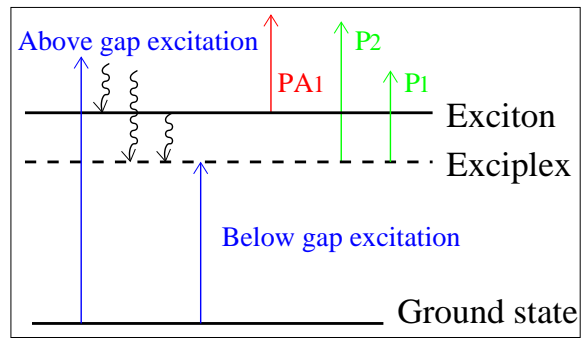}
\caption[a]{(Color online) Schematic of the consequences of above- and below-gap photoexcitations. Vertical straight arrows pointing up denote absorptions; curly arrows pointing down denote nonradiative relaxations.}
\label{schematic}
\end{figure}

 We have already presented the implications of our results for ground state absorption and PL in the previous section. Our determination that the lowest exciplex in PPV-C$_{60}$ is nearly completely ionic agrees with the conclusion of Osikowicz {\it et al.} \cite{Osikowicz07a} Our conclusion that there occur multiple exciplexes in PPV-C$_{60}$ agrees with the conclusion of Drori {\it et al.} \cite{Drori08a} Our interpretations of the transient absorptions that result from above- and below-gap excitations \cite{Drori08a} are shown schematically in Fig.~\ref{schematic}. The figure is not arrived at from the theoretical calculations alone as the rates of various intermediate nonradiative processes indicated in the figure cannot be obtained theoretically. Our conclusions are rather reached by combining arguments that follow from experiments as well as the calculations reported in the previous section.
\par The high pump energy (3.1 eV) of the above-gap excitation \cite{Drori08a} can take the system to the continuum band of PPV, \cite{Chandross97a} albeit with small cross-section, generating some mobile polarons directly. In addition, absorption can also occur to the lowest C$_{60a}^*$ states, which are at this energy and are weakly dipole coupled to the ground state (strong molecular absorptions in C$_{60}$ occur at much higher energy). In either case, there is rapid nonradiative decay to the lower-energy PPV$^*$C$_{60}$ exciton and the exciplex. The rate of this ultrafast decay cannot be estimated theoretically but it is easily seen from considerations of energy that no other explanation of the experiments is possible. Recall that all three PA bands P$_1$, PA$_1$, and P$_2$ are seen at zero time delay. \cite{Drori08a} The PA band PA$_1$ has previously been seen in solutions  as well as films of pure PPV derivatives \cite{Sheng07a} and has also been theoretically analyzed as the excitation from the PPV optical exciton to the mA$_g$. \cite{Chandross97a,Wang08a,Psiachos09a} We have not found any PA at this energy from PPVC$_{60a}^*$ or PPVC$_{60f}^*$ states. Considering all of the above PA$_1$ can only be an excitation from the PPV$^*$C$_{60}$ exciton. In contrast, PA bands P$_1$ and P$_2$ are never seen in solutions,  and hence cannot involve excitations from the neutral PPV chain. As mentioned above, we see complete absence of PAs from the C$_{60}^-$ component of the exciplex. The only possibilities are then that these are free polaron absorptions or absorptions from the PPV$^+$ component of the exciplex state. The similarities in the P$_1$ and P$_2$ bands obtained from above-gap and below-gap excitations, observed experimentally, show that these are absorptions from the exciplex. The absence of the PA$_1$ band in the above-gap PA at later times \cite{Drori08a} is therefore an unambiguous signature that PA now is from the exciplex alone, which, as seen in Figs.~5, does not absorb at energy PA$_1$.
\par PA resulting from below-gap excitation is also from the same exciplex. We have already pointed out that the scaled calculated energy of the exciplex is close to that determined by Drori {\it et al.} \cite{Drori08a} As in the case of single-component PCPs, PA bands P$_1$ and P$_2$ originate from a bound state. This explains the immobility of the ``polarons'' noted by Drori {\it et al.}, \cite{Drori08a} especially upon below-gap excitation. The IRAV accompanying the PAs P$_1$ and P$_2$ is due to the asymmetric nature of the exciplex wave function. Unlike in the CTX in pure MEH-PPV, the PPV molecule in the exciplex has a definite charge (positive) and the coupling between the electronic and vibrational modes is nearly the same as in an isolated chain with a localized defect. While it is not clear whether the charge transfer PA-located energetically between P$_1$ and PA$_1$ is seen experimentally in the blend of MEH-PPV and C$_{60}$, \cite{Drori08a} we point out that Holt {\it et al.} claim to have observed precisely this absorption in a blend of MEH-PPV and acceptor 2,4,7-trinitro-9-fluorenone (TNF) (see Fig.~5 in Ref. 27). TNF is a stronger acceptor than C$_{60}$, and hence leads to even larger HOMO-LUMO offsets than in PPV-C$_{60}$. Larger offsets in turn enhance the CT energy further, thereby probably leading to a larger splitting between P$_1$ and CT, which is necessary to observe a distinct CT band in PA.
\par In summary, ground state optical absorptions, as well as ps PAs in blends of PPV derivatives and fullerenes can be understood within the standard PPP model, with the intermolecular dielectric screening $\kappa_{\perp}$ as an adjustable variable. PA experiments find complete absence of the PA$_1$ band upon below-gap excitation, indicating nearly completely ionic character of the lowest exciplex. We find nearly completely ionic lowest exciplex for many different relative orientations between the PPV and C$_{60}$ molecules, varying $\kappa_{\perp}$ very slightly. The relative orientations of the PPV oligomer and C$_{60}$ have minor consequences on the overall energy locations of the PA bands. We emphasize that in contradiction to some recent proposals, the photophysics of the blend cannot be understood in the context of  ground-state charge transfer. Mulliken's original theory of ground state charge transfer \cite{Mulliken52a} assumed that the intramolecular excitations are very high in energy and the donor-acceptor complex can be described using only two basis functions, neutral DA and ionic D$^+$A$^-$. This two-state approximation breaks down when either or both the optical excitations D$^*$A and DA$^*$ are close in energy to D$^+$A$^-$, as is true in complexes involving PCPs. The CTX in pure PCPs and the exciplex in blends both arise predominantly from excited-state charge transfer, involving D$^+$A$^-$ and the intramolecular excitations. The key difference between the CTX and the exciplex is that the former is predominantly neutral and the latter is ionic. We have found in both single-component systems and blends a charge transfer PA distinct from P$_1$ and PA$_1$. Whether this PA will be seen in systems other than MEH-PPV:TNF is of interest. Also of interest are the high-energy exciplex states of Fig.\ref{eng-spctrm}(a) with lower ionicities and hence stronger electronic couplings with the optical exciton in the PCP. These are also expected to be electronically coupled to extended states involving more distant molecules in a multilayer system. \cite{Muntwiler08a} We are currently investigating model multilayer systems theoretically.
\begin{figure}
\includegraphics[width=3.3in]{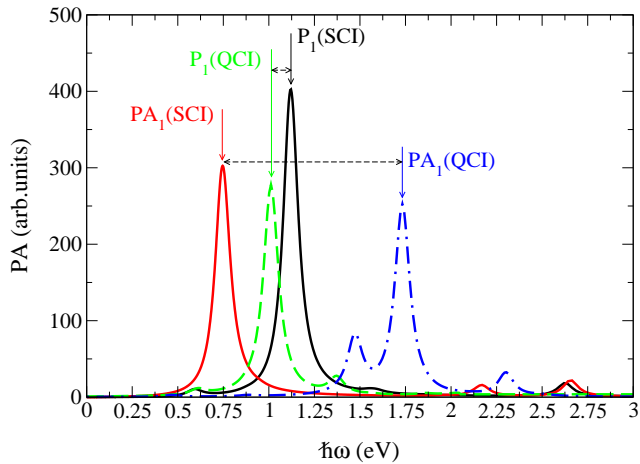}
\caption[a]{(Color online) Relative locations of the P$_{1}$ and PA$_{1}$ bands, calculated using SCI and QCI.} 
\label{appdx1}
\end{figure}
\section*{ACKNOWLEDGMENTS}
\par We are grateful to A. Shukla and Z. Wang for help with computations, and to Z. V. Vardeny and L. J. Rothberg  for helpful discussions. This work was supported by NSF under Grant No. DMR-0705163.
\section*{APPENDIX: HIGHER-ORDER CI EFFECTS ON THE PA LOCATIONS}
\par We report here SCI and QCI calculations of PA for an artificial DA complex consisting of a pair of PPV3 oligomers, one of which can act as a donor in the excited-state, while the other behaves as an acceptor. We achieve this by adding a term ${\sum_{\mu,i}(-1)^{\mu} \epsilon n_{\mu,i}}$ to $H_{intra}$ in Eq.~\ref{H_intra}. We assume a perfectly stacked cofacial dimer, as in our earlier work for single-component systems, \cite{Wang08a,Psiachos09a} and limit $t_{ij}^{\perp}$ = 0.1 eV to nearest interchain neighbors (note that the difference between SCI and QCI arises from the Coulomb interactions alone and $t_{ij}^{\perp}$ plays a weak role). As with the PPV-C$_{60}$ calculations with geometry I, we choose an intermolecular distance of 0.4 nm and $\kappa_{\perp}=1.3$. With the total number of carbon atoms $N=44$, QCI is feasible; the dimension of the QCI Hamiltonian matrix is 1833276. With $\epsilon=$0.185 eV the energy offset between the Hartree-Fock HOMO levels is 0.37 eV, equal to the offset between the HOMOs of PPV and C$_{60}$ in Fig.\ref{HFMO}. With these parameter sets, the correlated energy spectrum near the optical edge is very similar to that in PPV-C$_{60}$: the lowest excited-state is the exciplex with ionicities of 0.95 with SCI and 0.93 with QCI, and the optical exciton is 0.88 eV above the exciplex. Meaningful comparisons of PA calculations for the two cases can therefore be done.
\par In Fig.\ref{appdx1}, we show the calculated PA bands from the exciplex and exciton, using both SCI and QCI. The final states of PA bands labeled P$_1$ from the exciplex have ionicities 0.91 and 0.80, within SCI and QCI, respectively. Thus both the initial and final states here are predominantly ionic. The PA bands from the exciton, PA$_{1}$, in contrast, have final-state ionicities of 0 and 0.28 within SCI and QCI, respectively. As expected the initial and final states here are predominantly covalent. As seen in Fig.\ref{appdx1}, P$_{1}$ and PA$_{1}$ bands within SCI occur at energies $\sim$ 1.12 and 0.74 eV, respectively. Their proximity is similar to that in PPV-C$_{60}$ in Fig.~\ref{PA-absorption}(a), with the slightly higher energy of P$_1$ here ascribed to the large intraband gaps in the short PPV3 oligomer. As Fig.\ref{appdx1} also shows, the energies of the PA bands originating from the exciplex show relatively modest shifts between SCI and QCI. In contrast, PA originating from the exciton is affected very strongly. Thus while P$_{1}$ red-shifts weakly from 1.12 to 1.01 eV, PA$_{1}$ displays an enormous blue-shift from 0.74 to 1.73 eV, such that it now appears considerably above P$_{1}$. As claimed in Sec.\ref{sec:results}, these results are actually to be anticipated from earlier calculations on the nature of the mA$_g$ two-photon state, and provide the justification for our rigidly blue-shifting the calculated PA$_{1}$ band of PPV-C$_{60}$ in Fig.\ref{PA-absorption}(b) while leaving the energy location of P$_{1}$ intact.

\end{document}